\documentclass[amsmath,amssymb,numerical,super,comma, sort&compress, aip,jcp,16pt]{revtex4-1}

\usepackage{mathptmx}
\usepackage[scaled=0.90]{helvet}
\usepackage{courier}
\normalfont

\usepackage{graphicx}
\usepackage{braket}
\usepackage{color}
\usepackage{simplewick}
\usepackage{mathptmx}
\usepackage{booktabs}

\renewcommand{\footnote}[1]{{\color{green}[#1]}}

\usepackage{cancel}
\usepackage{amsmath}
\usepackage{physics}
\usepackage{graphicx}
\usepackage{amssymb}
\usepackage{amsthm}
\usepackage{tabularx}
\usepackage{tikz}
\usetikzlibrary{arrows,shapes}
\usetikzlibrary{trees}
\usetikzlibrary{matrix,arrows} 				
\usetikzlibrary{positioning}				
\usetikzlibrary{calc,through}				
\usetikzlibrary{decorations.pathreplacing}  
\usepackage{pgffor}							

\usetikzlibrary{decorations.pathmorphing}	
\usetikzlibrary{decorations.markings}
\tikzset{
    vector/.style={decorate, decoration={snake}, draw},
	provector/.style={decorate, decoration={snake,amplitude=2.5pt}, draw},
	antivector/.style={decorate, decoration={snake,amplitude=-2.5pt}, draw},
    fermion/.style={draw=black, postaction={decorate},
        decoration={markings,mark=at position .55 with {\arrow[draw=black]{>}}}},
    fermionbar/.style={draw=black, postaction={decorate},
        decoration={markings,mark=at position .55 with {\arrow[draw=black]{<}}}},
    fermionnoarrow/.style={draw=black},
    gluon/.style={decorate, draw=black,
        decoration={coil,amplitude=4pt, segment length=5pt}},
    scalar/.style={dashed,draw=black, postaction={decorate},
        decoration={markings,mark=at position .55 with {\arrow[draw=black]{>}}}},
    scalarbar/.style={dashed,draw=black, postaction={decorate},
        decoration={markings,mark=at position .55 with {\arrow[draw=black]{<}}}},
    scalarnoarrow/.style={dashed,draw=black},
    electron/.style={draw=black, postaction={decorate},
        decoration={markings,mark=at position .55 with {\arrow[draw=black]{>}}}},
	bigvector/.style={decorate, decoration={snake,amplitude=4pt}, draw},
}

\tikzstyle{block} = [draw, rectangle,
    minimum height=3em, minimum width=6em]
\usepackage{caption}
\usepackage{subcaption}

\usepackage{mathtools}

\newtagform{Alph}[\renewcommand{\theequation}{\Alph{equation}}]()
\newtagform{roman}[\renewcommand{\theequation}{\roman{equation}}]()
\newtagform{scroman}[\renewcommand{\theequation}{\scshape\roman{equation}}][]

\begin{document}

\title{{\color{blue}A route to improving RPA excitation energies through its connection to equation-of-motion coupled cluster theory}\vspace*{0.18cm}}
\author{Varun Rishi}
\email{vrishi@caltech.edu}
\affiliation{Chemistry and Chemical Engineering, California Institute of Technology, Pasadena, California 91125, USA}
\author{Ajith Perera}
\email{perera@qtp.ufl.edu}
\affiliation{Quantum Theory Project, University of Florida, Gainesville, Florida 32611, USA}
\author{Rodney J. Bartlett}
\email{bartlett@qtp.ufl.edu}
\affiliation{Quantum Theory Project, University of Florida, Gainesville, Florida 32611, USA}

\date{\today}

\begin{abstract}
We revisit the connection between equation-of-motion coupled cluster (EOM-CC) and random phase approximation (RPA) explored recently by Berkelbach {[J. Chem. Phys. \textbf{149}, 041103 (2018)]} and unify various methodological aspects of these diverse treatment of ground and excited states. The identity of RPA and EOM-CC based on the ring coupled cluster doubles is established with numerical results which was proved previously on theoretical grounds. We then introduce new approximations in EOM-CC and RPA family of methods, assess their numerical performance and explore a way to reap the benefits of such a connection to improve on excitation energies. Our results suggest that addition of perturbative corrections to account for double excitations and missing exchange effects could result in significantly improved estimates.
\end{abstract}

\maketitle

\section{Introduction}


The random phase approximation (RPA),\cite{Pines1951,Pines1952, Pines1953} in its various flavors, holds a historical place in the development of theoretical models in condensed matter physics and quantum chemistry \cite{Furche2017, Furche2012, Scheffler2012, Hesselmann2011, rowe2010}. From the viewpoint of quantum chemistry RPA offers a dilemma. RPA’s primary  role in quantum chemistry before its recent resurgence through use in density functional theory (DFT),\cite{Furche2001} had been to describe approximations to excitation energies of molecules as an effective one-particle theory. The dual role of RPA, both as a ground state and excited method, has lead to a longstanding debate whether RPA, a correlated method for ground state, is a also correlated for excited states. Coupled cluster (CC) or configuration interaction (CI) theory focuses on the correlation problem for the ground state and its subsequent effect on all other properties including excited states. In many-body physics the term RPA corresponds to the well-known infinite sum of ring diagrams\cite{Brandow1967,Freeman1977} that describes the correlation energy of the high-density electron gas.\cite{Pines1951, Pines1952, Pines1953}. An equivalence has been known for the ground state energy for RPA and a particular approximation to coupled cluster doubles (ring CCD) method\cite{Freeman1977, Scuseria2008, Jansen2010}. This equivalence was also to shown exist when RPA is derived as an excited state method using the equation of motion (EOM-CC) approach limited to single excitations starting from the corresponding ring-CCD ground state.\cite{Berkelbach2018} This contradicts the origin of correlation in single-excitation (and de-excitation) based RPA theory and raises pertinent doubts about the role of the ground state therein.

The above questions have caused much consternation over the years, as it was felt that an identification of a \textit{consistent} correlated ground state\cite{Ellis1970,Ostlund1971,Ohrn1979,Catara1996} could be used to define an optimum method to build electron correlation into RPA’s excited states, enabling it to provide much more accurate results than the normal RPA based on an assumption of HF ground state. One initial attempt to introduce a correlated ground state was done by Shibuya and McKoy, leading to what is termed higher-RPA.\cite{Mckoy1970, Mckoy1970b, Mckoy1971} By using a first-order correlated reference wavefunction (MBPT(1)) as the correlated ground state in the RPA evaluation, the numerical excitation energies limited to single excitations were shown to be improved. In principle, this approach could be generalized to higher-orders in MBPT, or even to CC theory. But even then assessing the comparative importance of ground state correlation and  more extensive intermediate states (such as $2p2h$, $2h2p$ and further) remains difficult. Another such approach was the second-order polarization propagator approximation (SOPPA)  of Oddershede, Sauer and co-workers,\cite{Oddershede1980, Oddershede1996, Sauer2020} which is an independent route to identify on formal grounds the kind of wavefunction that would offer a consistent RPA ground state. This led to the introduction of the anti-symmetrized geminal power (AGP) wavefunction of Linderberg, \"{O}hrn, Weiner and Goscinski, equivalent to a projected BCS solution.\cite{Ohrn1979, Weiner1980, Goscinski1981} A new route could be suggested, too. Verma and Bartlett show that a one-particle correlation potential generated from RPA’s ground state correlation can be rigorously defined to augment the usual Fock operator of standard RPA, which would incorporate correlation effects in RPA in a different way, more familiar in TDDFT circles.\cite{Verma2012} But unlike others in TDDFT, this RPA-OEP potential is \textit{ab-initio}. This can also be done for CC approximations in the same way.

\begin{figure}[h!tbp]
\centering
\includegraphics[width=\textwidth]{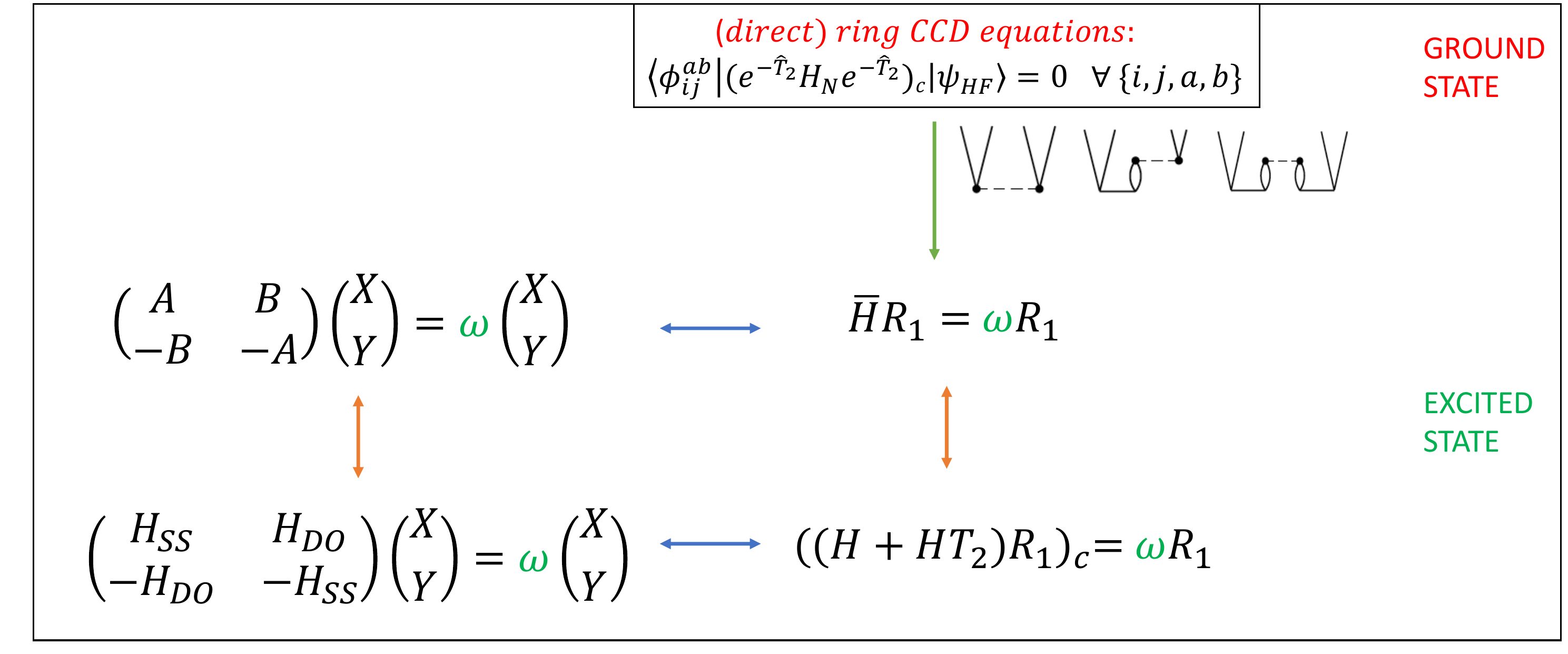}
\caption{Relationship between random phase approximation for excited states and the equation of motion coupled cluster based on (direct) ring CCD} \label{Relation_EOMCC_RPA}
\end{figure}

In context of this work, we are interested in the ramifications of the equivalence of RPA with ring CCD, with and without exchange contributions,\cite{Freeman1977, Scuseria2008, Angyan2010} and the subsequent equivalence for excitation energies predicted by RPA and the equation-of-motion approach based on (ring) coupled cluster doubles (rCCD) ground state wavefunction as presented recently by Berkelbach\cite{Berkelbach2018}. This identity has a potential to enrich both the RPA and EOM-CC family of methods (Fig \ref{Relation_EOMCC_RPA}). In particular, we could benefit from the past two decades of work introducing and establishing EOM-CC methods as the an accurate benchmark for excited and ionized states of molecular systems.

Equation of motion approach based on a coupled cluster\cite{Cizek66,shavitt2009} ground state reference  (EOM-CC)\cite{sekino1984, geertsen1989,comeau1993, stanton1993} has been shown to accurately calculate the excitation energies for molecular systems\cite{Watts2008,Bartlett2012,krylov2008,Izsak2019} and, more recently, for condensed matter systems\cite{mcclain2017, XiaoWang2020}. EOM-CCSD method ($ N^{6}$ scaling) offers an accuracy of 0.2-0.3 eV for states with dominant single excitation character\cite{watson2013, Rishi2014} and has recently been applied to large molecular systems based on local coupled cluster schemes\cite{Dutta2016}. Complete inclusion of triples leads to EOM-CCSDT method\cite{kucharski2001, Musial2003, Musial2004} whose error estimate was shown to be 0.05-0.1 eV compared to the full configuration interaction (FCI) results for singly and doubly excited states. Several approaches with approximate treatment of triples exist (EOM-CCSDT-3\cite{watts1994}, EOM-CCSD(T)\cite{watts1995, watson2013}, EOM-CCSDR3,\cite{Christiansen1996b} EOM-CCSDT-1a*\cite{Matthews2016} and CC3\cite{Christiansen1998b}) and offer a balance between accuracy and computational cost\cite{watts1994, watson2013,watts1995, watts1994, christiansen1995}. But even EOM-CCSD is computationally expensive for moderately large systems\cite{Izsak2019, tripathi2019} and hence, approximations (EOM MBPT(2)/EOM CCSD(2), partitioned EOM-MBPT(2), CIS(D) and CIS(D$_\infty$), potentially scaling lower than $N^{6}$, have been developed\cite{gwaltney1996, headgordon1994, goings2014,  Tajti2016, Tajti2018, Tajti2019, Kumar2017,Rishi2015,Boguslawski2016, Park2018, Besley2019, wang2019, Dutta2016, peng2018}. In the context of this and future studies, we are interested in the potential for the development of a low-scaling EOM-CC method inspired by its RPA connection.

As the excitation energy estimates from RPA show large deviation from highly accurate EOM-CCSD, an effort in the direction of improving it by the inclusion of double excitation effects, RPA(D), was made by Sauer \textit{et al.}\cite{Sauer1998, Sauer2020} This is done in similar spirit to the non-iterative (D) correction to CIS method as put forth by Head-Gordon \textit{et al.}\cite{headgordon1994} In the realm of polarization propagator approaches, RPA is seen to incorporate first-order effects and consequently, second-order approaches (SOPPA) have been proposed as well\cite{Oddershede1980,sauer1997}. In a recent work using RPA based on KS states, a TDDFT formulation was applied to treat molecular excited states\cite{Hesselmann2015}. In a different direction, exploratory work by Pernal \textit{et al.},\cite{Pernal2012, Pernal2018} DePrince \textit{et al.},\cite{ Deprince2016, Deprince2018} and Ayers \textit{et al.}\cite{VanAggelen2013} studied the performance of extended RPA (ERPA) approach, using geminal wavefunction as a reference in a formalism employing reduced density matrices (RDMs). A formulation of particle-particle RPA, analogous to the more explored particle-hole RPA, has also been used to calculate excited state energies (double excitations) and other properties.\cite{Yang2013b,Yang2013c, Yang2014} 

Recent developments enriching the RPA toolbox also include devising of F12 corrections by Klopper \textit{et al.},\cite{Hehn2013, Hehn2015, HEHN2016} a multi-reference formulation,\cite{Szabados2017, Szabados2020} addition of the effect of orbital relaxation,\cite{Ren2011} analytical gradients,\cite{Furche2014, Mussard2014, Ochsenfeld2018} and exchange corrections to direct RPA. RPA is non-perturbative and hence, more robust when dealing with state degeneracies.\cite{Fuchs2005} As dRPA suffers from self-interaction error (it violates Pauli's exclusion principle), there have been proposals to eliminate one-electron self-interaction error and reduce the many-body self-interaction error as well.\cite{Hummel2019} The acronyms SOX, SOSEX\cite{Gruneis2009}, RPA+, IOSEX\cite{Kresse2016} and more recently gRPA+\cite{Gould2019} all refer to such approximations. Another way to overcome Pauli principle violation was reported by Kosov\cite{Kosov2017} who proposed an a posteriori correction to single particle density matrix obtained after HF based dRPA calculation. Unresolved connections exist between the treatment of static correlation and the presence of self-interaction error.\cite{Henderson2010} rCCD methods have also been studied and extended through the paradigm of coupled cluster perturbation theory where rCCD is considered to be the zeroth-order problem.\cite{Lotrich2011} Another stream of work has focused on reducing the computational cost associated with RPA methods.\cite{Gonze2002, Furche2008} Formally RPA scales as $N^{6}$, but many recent work employing density-fitted (DF) basis have reported cubic scaling ($N^{3}$) implementations thus leading to massive reduction in cost.\cite{Furche2010b, Hutter2016, Kresse2014, Ochsenfeld2016, Ochsenfeld2019} Employing localized orbitals has led to linear scaling implementation\cite{Kallay2015b} for dRPA and has resulted in lower scaling for models including exchange corrections.\cite{Hesselmann2017} Most of these developments have focused on ground state treatment but they have important consequences for excited state formulation as well. 


The intent of this paper is to explore the systematic inclusion of electron correlation in ground state wavefunction via coupled-cluster theory to obtain improved approximations to RPA. Compared to propagator methods where the correlated reference state is considered secondary (an alternate view could be seen in the work of Ortiz, Sauer and Dreuw and co-workers),\cite{Ortiz1992, sauer1997, Hodecker2019, Hodecker2020} using the CC wavefunction as the reference state can have some major consequences. For example, in the electron propagator (EP) once coupled cluster wavefunction for ground state ($\ket{\Psi_{CC}}=e^{T}\ket{\Psi_{HF}}$) used, several benefits occur.\cite{Meissner1993, Nooijen1992, Nooijen1993} First, the frequency dependence of the self-energy is rigorously removed. Second, the EP divides naturally into an ionization problem, IP-EOM-CC and an independent electron affinity problem, EA-EOM-CC, eliminating the unphysical coupling that exists in the EP.\cite{nooijen1995} Third, the dominant IP’s and EA’s are the first solutions obtained from diagonalizing the similarity transformed Hamiltonian ($\bar{H}$ in CC theory) in the IP/EA-EOM-CC equations, instead of being  hidden among the full set of  all possible solutions of the EP  like those due to ‘shake-ups’. Somewhat analogously, this paper wants to address the insertion of the CC wavefunction as the correlated reference into the RPA excited state problem in general, or the polarization propagator in particular; and observe the consequences. All such approaches begin with the electronic excitation energy variant (EE)-EOM-CC,\cite{sekino1984, comeau1993, stanton1993} where electron correlation appears in: (1) similarity transformed Hamiltonian, $\bar{H} = e^{-T}He^{T}$, and (2) its projection onto an excitation only Hilbert space that defines the correlated excited states, \{$R_{k}$\} and their left-hand complements,  \{$L_{k}$\} and thus the $\bar{H}{R_{k}}=\omega_{k}{R_{k}}$, that provides the \{$\omega_{k}$\} with their bi-orthogonal norm, $\bra{L_{k}}\ket{R_k}=\delta_{kl}$ and their associated transition moments. The full CI requires that the Hilbert space generated by $T$ and used in $\bar{H}$ include the Fermi vacuum and all possible excitations from it. This is a formally important boundary condition, but would seem to be redundant if the goal is to find a mutually beneficial level of ground state correlation coupled to a representation of the excited states that would maximize accuracy while offering a reasonable computational scaling. We develop approximations based on EOM-CC approach, starting from those numerically equivalent to (d)RPA, and assess the suitability of various iterative and perturbative corrections to these parent models by a numerical analysis of the results for small molecular systems.


The manuscript is organized as follows. In the next section (\ref{Theory}), we begin by reiterating how a connection of RPA with the coupled cluster approximation for ground state leads to a relationship between RPA excited states and EOM based on CC ground state and consequently, how different corrections to RPA excitation energy could potentially improve it. We analyze the results of these approximations in the subsequent section (\ref{Results}).

\section{Theory}\label{Theory}

\subsection{Coupled cluster connection to the correlated ground state for RPA}

We briefly discuss the theory of random phase approximation and its connection to the coupled cluster methodology. Denoting the ground state Hartree-Fock (HF) wavefunction as $ \ket{\Psi_{HF}}$, the RPA creation operator, composed of excitation(particle-hole creation) and de-excitation(particle-hole annihilation) operators, is 
\begin{equation}
O^{\dagger} = X_{ai}\{ a_{a}^{\dagger}a_{i}\} - Y_{ia}\{a_{i}^{\dagger}a_{a} \}     \label{RPA_excitation_operator}
\end{equation}
The fact that RPA allows for ground state correlation through the de-excitation operators suggests that the corresponding wave function, $ \ket{\Psi_{RPA}}$, is not adequately represented by a single Slater determinant. Associated with this concern is the fulfillment of the killer condition i.e. the application of RPA annihilation operator should destroy the ground state
\begin{equation}
O{\ket{\Psi_{RPA}}} = 0  
\end{equation}
If the RPA ground state is chosen to be a HF wavefunction, the killer condition is not satisfied 
\begin{equation}
O{\ket{\Psi_{HF}}} \neq 0  
\end{equation}
There has been an unending debate on the true nature of RPA ground state and a correlated ground state may satisfy this condition.\cite{Ohrn1979, Schuck2013} The work of Scuseria \textit{et al.}\cite{Scuseria2008} showed analytical equivalence between a subset of CCD (named direct ring-CCD or drCCD) residual equations and direct random phase approximation (dRPA) equations. This potentially provides a consistent ground state for RPA approximation but the wavefunction form still appears to be unclear as there is no clear exponential form deducible from the truncated subset of drCCD residual equations. We briefly introduce these equations.  
Considering the normal-ordered Hamiltonian as  
\begin{equation}
\hat{H} = {\hat{f}_{pq}}\{a_{p}^{+}a_{q}\} + \dfrac{1}{4} {\bra{pq}}{\ket{rs}}\{a_{p}^{\dagger}a_{q}^{\dagger}a_{s}a_{r}\}
\end{equation}
where $p,q,\dots $  are spin-orbitals that are obtained after convergence of HF equations: $ \hat{f}\ket{p} = \epsilon_{p}\ket{p}$. Antisymmetrized integrals are defined as ${\bra{pq}}{\ket{rs}} = {\bra{pq}}{{rs}\rangle} - {\bra{pq}}{{sr}\rangle}$, and ${\bra{p}}{{q} \rangle} = \delta_{pq}$. Einstein summation is assumed wherever required in the rest of the manuscript.  The coupled cluster doubles parameterization of the wavefunction is written as,
\begin{equation}
\ket{\Psi_{CCD}} = e^{T_{2}}\ket{\Psi_{HF}}    
\end{equation}
The double excitation operator, $T_{2} = \dfrac{1}{4}{t_{ij}^{ab}}\{a_{a}^{\dagger}a_{b}^{\dagger}{a_j}{a_i}\}$ is defined as such and the associated amplitudes ${t_{ij}^{ab}}$ could be calculated from the doubles residual equation,  
\begin{equation}
R_{ij}^{ab} = \bra{D}\bar{H}\ket{\Psi_{HF}} = 0    
\end{equation}
where $\bar{H} = e^{-T}{H}e^{T}$ is the \textit{similarity transformed Hamiltonian} and the projection from left is through doubly excited determinants ($D$). Without delving into the complete derivation of coupled cluster doubles (CCD) equations,\cite{Cizek66, bartlett1978, pople1978, shavitt2009} we would like to draw connections to the terms that are present in the direct-ring (without exchange) and ring CCD (with exchange) approximation. To revisit the ring CCD equations for ground state, we have 
\begin{equation}
R_{ij}^{ab} = {\bra{ab}}{\ket{ij}} + (\epsilon_{a} + \epsilon_{b} - \epsilon_{i} - \epsilon_{j})t_{ij}^{ab} + P(ij)P(ab)t_{im}^{ae} W_{ej}^{mb}   
\end{equation}
Only the Coulomb integrals are used in case of direct ring CCD equations and the consequent $t_{ij}^{ab}$ amplitudes are not antisymmetric with the permutation of indices ($i$ with $j$ and $a$ with $b$). We also restrict the permutation operator ($ P(ij)P(ab)$),  that antisymmetrizes the third term in the residual, to keep only the contributions that that are totally symmetric to the simultaneous exchange of indices $i$ with $j$ and $a$ with $b$. That is $P(ij)P(ab) = 1 + P({ij}) P(ab)$ and henceforth simply designated as $P_{-}(ij)P_{-}{(ab)}$. This leads to a form of rCCD equation given as, 
\begin{equation}
R_{ij}^{ab} = {\bra{ab}}{\ket{ij}} + (\epsilon_{a} + \epsilon_{b} - \epsilon_{i} - \epsilon_{j}) t_{ij}^{ab} + ( t_{im}^{ae} {\bra{mb}}{\ket{ej}} +  t_{jm}^{be} {\bra{ma}}{\ket{ei}} ) + {t_{im}^{ae} {\bra{mn}}{\ket{ef}}} {t_{nj}^{bf}}
\label{CCD}
\end{equation}
The amplitudes obtained are no longer antisymmetric in drCCD or rCCD. Restated, this is the Pauli principle violation not observed in CC methods or in the many CCD/CCSD like approximations\cite{Paldus2017} with two exceptions: distinguishable cluster (DCD, DCSD and DCSDT)\cite{Kats2013, Rishi2016, Kats2019, Rishi2017, Rishi2019, Rishi2018} and approximate coupled pair method (ACP-D14)\cite{Paldus1980, Piecuch1991}. The consequences of this violation, which is a form of many-body self-interaction, is unclear. A link between self-interaction error present in RPA as used in density functional theory (DFT) and the ability to capture static correlation has been mentioned by Henderson \textit{et al}.\cite{Henderson2010}

It is important to point out that as opposed to the CCD wave function, neither of the rCCD or drCCD wave functions can be written in terms of an exponential parameterization. It might be fair to say that these models do not have compact analytic form of the wavefunction. The equations for the amplitudes (and for the corresponding energy) come out to be a subset of CCD equations but cannot be traced back to an exponential wavefunction ansatz themselves. This distinction to CCD is extensively used in subsequent development in this work. The forms of the energy expressions are discussed in detail in appendix. Let us now show the identities of rCCD and drCCD to RPA and dRPA respectively. Following Scuseria \textit{et al.},\cite{Scuseria2008} we introduce the terms $A$ and $B$ defined as $B_{ia,bj}=\langle{ab}||{ij}\rangle$ and $A_{ia,bj}=(\epsilon_{a}-\epsilon_{i} ) \delta_{ab} \delta_{ij} + \langle{ia}||{bj}\rangle $ in the equation (\ref{CCD}) and with some manipulations, we obtain
\begin{align}
R_{ij}^{ab} = {\bra{ab}}{\ket{ij}} +  t_{im}^{ae} <mb||ej> + (\epsilon_{e} - \epsilon_{m})\delta_{eb} \delta_{mj}t_{im}^{ae} + t_{jm}^{be} {\bra{ma}}{\ket{ei}}  + (\epsilon_{e} - \epsilon_{m})\delta_{ea} \delta_{mi} t_{jm}^{be} \\ \notag
+  {t_{im}^{ae}}{\bra{mn}}{\ket{ef}} {t_{nj}^{bf}}
\end{align}
Defining $T_{ai,bj} = {t_{ij}^{ab}}$ and rewriting the equation,  we obtain
\begin{equation}
B + TA + AT + TBT  = 0\label{Ricatti}
\end{equation}
The equation (\ref{Ricatti}) is identical to the expression obtained from the RPA eigenvalue equation,
\begin{equation}
\begin{pmatrix}
A & B \\  -B & -A
\end{pmatrix}
\begin{pmatrix}
X \\ Y
\end{pmatrix}
= 
\begin{pmatrix}
X \\ Y
\end{pmatrix}\omega
\label{RPAeigen}
\end{equation}
where $A$ and $B$ are defined as
\begin{equation}
A= \langle{\Phi_{i}^{a}}|{H}|{\Phi_{j}^{b}}\rangle
\end{equation}
\begin{equation}
B= \langle{\Phi_{ij}^{ab}}|{H}|{\Psi_{HF}}\rangle
\end{equation}
and $\omega$ is the diagonal matrix containing the single excitation and de-excitation energies. We obtain an identical set of excitation energies and de-excitation energies. Depending on whether the above matrix is positive semi-definite, the eigenvalues may or may not be real. For the case of direct RPA, it is always positive definite and eigenvalues are real. This is not guaranteed for RPA (which includes exchange terms). The diagonalization of the matrix yields excitation energies ($\omega$) for singly excited states and the eigenvectors which are the coefficients in terms of single excitations and de-excitations. 

Recent work by Berkelbach\cite{Berkelbach2018} shows the numerical equivalence of the RPA excitation energies with those obtained from diagonalization of similarity transformed Hamiltonian in the space of single excitation determinants, $\bar{H}_{SS}$, constructed for equation-of-motion approach based on ring CCD (EOM rCCD),
\[
\bar{H}_{SS}=
  \begin{bmatrix}
   <S|\bar{H}|S> \\
  \end{bmatrix}
\]
As rCCD and drCCD models are approximate ground state CC methods, we should be able to follow EOM-CC formalism to develop corresponding excited state analogs. Also, as stated earlier, the rCCD and drCCD only have defining equations for the amplitudes and for the corresponding energy, and unlike CCD that they do not conform to an exponential parameterization. Nevertheless, all the density matrices, and therefore the energy and all the other molecular properties can be obtained since having a compact analytic expression for the wave function is not a prerequisite to formulate the density matrices. We proceed in formulation of excited state method based on rCCD ground state by determining the corresponding effective Hamiltonian.


\subsection{Towards excited states through the formulation of $\Lambda$ equations and effective Hamiltonian ($\bar{H}$) for ring-CCD}

\begin{figure}[h!tbp]
\centering
\includegraphics[scale=0.7]{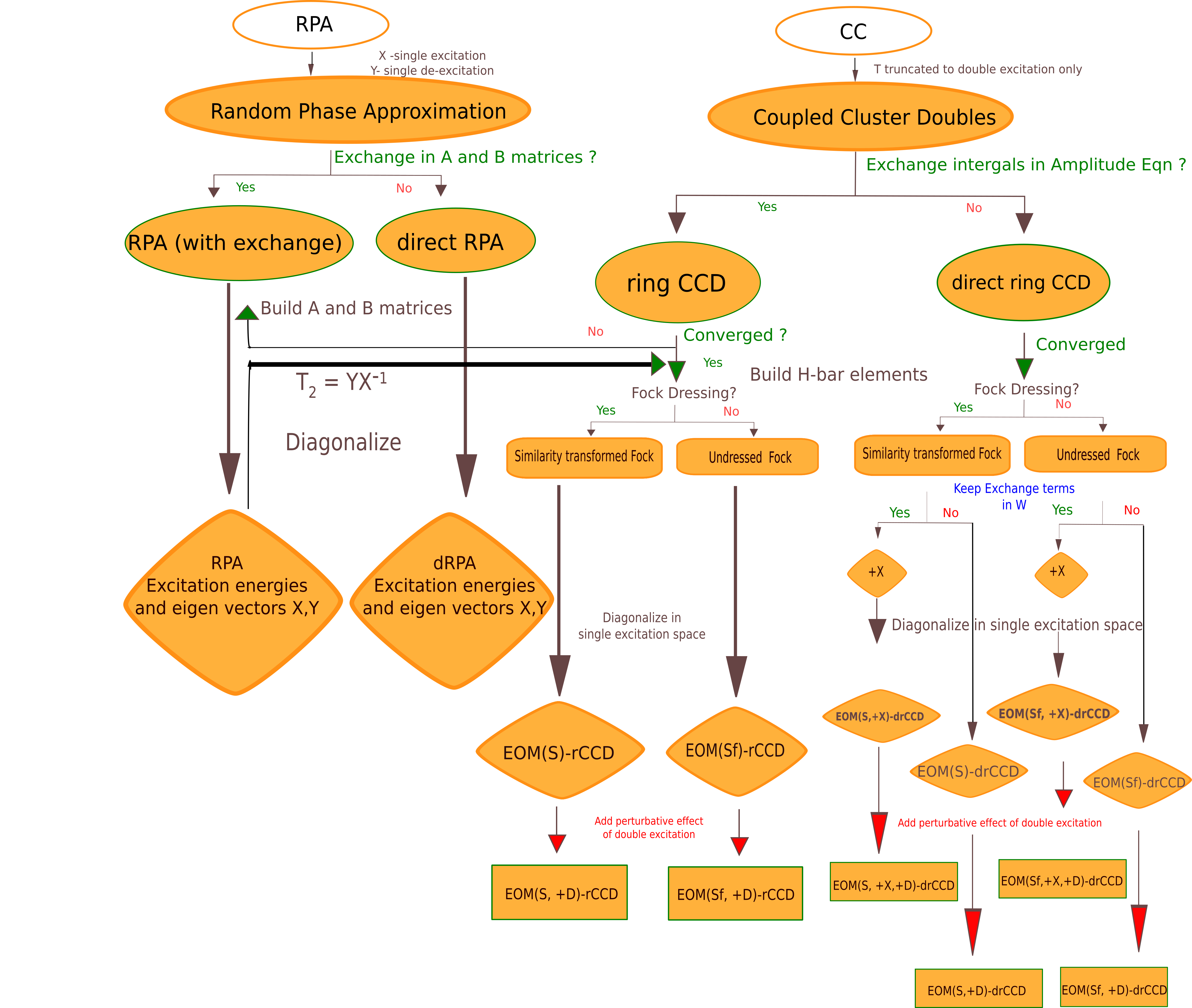}
\caption{A flowchart depicting various methods in CC and RPA family studied in this paper} \label{Flowchart_EOMCC_RPA}
\end{figure}

CC is a non-symmetric theory. As a result, the left and right hand wave functions are not identical and the left hand wave function must be determined independently. Both the rCCD and drCCD, which are approximations to CCD, share this property. As we have an equation that defines the $T_{2}$ excitation amplitudes for $\ket{\Psi_{CCD}}$, the Lagrangian multiplier technique is used to define an equation for $\Lambda_{2}$ de-excitation amplitudes defining $\bra{\Psi_{CCD}}$,
\begin{equation}
L= \bra{0}({H}{T_2})_c\ket{0} + \sum_{i,j,a,b} {\lambda_{ab}^{ij} R_{ij}^{ab}}
\end{equation}
We rewrite the (d)rCCD Lagrangian by simplifying the first-term (energy) and substituting for  $R_{ij}^{ab}$, 
\begin{equation}
L= \dfrac{1}{4} {\bra{ab}}{\ket{ij}} t_{ij}^{ab}+\lambda_{ab}^{ij} [{\bra{ab}}{\ket{ij}}+ t_{ij}^{ab} (f_{ae} \delta_{ea}+f_{bf} \delta_{fb} - f_{mi} \delta_{im} - f_{nj} \delta_{jn} )+ (1+P(ij)P(ab))t_{im}^{ae} W_{ej}^{mb} ]
\label{lambda}
\end{equation}
where $T_2$ satisfies the rCCD or drCCD residual equation and $\Lambda_{2}= \sum \lambda_{ab}^{ij} \{i^{\dagger}j^{\dagger}ba\}$ is defined to be a de-excitation operator corresponding to rCCD or drCCD. For comparison, the Lagrangian for the parent CCD model can be written compactly as
\begin{equation}
L=<0|\bar{H}|0> +\sum_{\mu}{ \lambda_{u} \bra{\mu}\bar{H}\ket{0}}>
\end{equation}
In order to obtain an equation for $\Lambda_{2}$,  we apply the stationary condition to $L$ with respect to variation of $T _2$ as shown in Eqn. (\ref{lambda}). Diagrammatically, this is identical to opening the $T _2$ and $H$ interacting lines in all
possible ways for the closed diagrams of the Langrangian.\cite{shavitt2009} This leads to,
\begin{align}
{\bra{ab}}{\ket{ij}}  + [(f_{ae} \delta_{ea}+f_{bf} \delta_{fb} - f_{mi}\delta_{im} -f_{nj} \delta_{jn} ) \lambda_{ab}^{ij} + (1+P(ij)P(ab)(W_{ej}^{mb} + \dfrac{1}{2}t_{nj}^{fb} \\ \notag
+  {\bra{mn}}{\ket{ef}})  \lambda_{ae}^{im}] =0
\end{align}
which is the \textit{lambda equation} for rCCD or drCCD.  This can be rewritten as,
\begin{equation}
\bar{H}_{ab}^{ij}+ \lambda_{eb}^{ij} (1+P (ab)) \bar{H}_{e}^{a}+\lambda_{ab}^{im} (1+ P(ij) \bar{H}_{m}^{i} + \lambda_{ae}^{im}  (1+P(ij)P(ab))\bar{H}_{ej}^{mb}=0
\end{equation}
where we have defined  $ \bar{H}_{ab}^{ij} ={\bra{ij}}{\ket{ab}}$, $  \bar{H}_{e}^{a} =f_{ae}$ , $\bar{H}_{m}^{i}=f_{mi}$ and $\bar{H}_{ej}^{mb}=W_{ej}^{mb} +  \dfrac{1}{2} t_{fb}^{nj} {\bra{mn}}{\ket{ef}}$ analogous to the coupled cluster effective Hamiltonian ($\bar{H}$) defined above. The critical difference however is that the rCCD and drCCD do not have an exponential form and $\bar{H}^{(d)rCCD} $ does not correspond to the  $\bar{H} = e^{-T} H e^{T}$ form. As we have shown above, one has to resort to the CC Lagrangian to formulate $\bar{H}$ for rCCD and drCCD. With this protocol, the $\bar{H}$ is defined such that eigenvectors of the lowest (ground state) left solution is $\Lambda_{2}$. Therefore, it follows that for rCCD and drCCD, 
\begin{equation}
\bar{H}_{ij}^{ab} = 0
\end{equation}
\begin{equation}
\bra{0}{\bar{H}}\ket{0}=\Delta{E} 
\end{equation}
\begin{equation}
\bra{0}{(1+ \Lambda_{2})}\bar{H}\ket{0} = \Delta{E}
\end{equation}
where $\Delta{E}$ is the rCCD or drCCD correlation energy. This establishes the fact that we can define a CC like effective Hamiltonian for rCCD and drCCD such that when acted upon from left by $\bra{0}(1+\Lambda_{2} )$  and from right  by $\ket{0}$  yields the rCCD (or drCCD) ground state energies. The $\bar{H}$ matrix for EOM-CCSD method has the following structure,\cite{comeau1993, stanton1993}
\begin{equation} 
\begin{pmatrix}
\Delta{E} & \bar{H}_{0S} & \bar{H}_{0D} \\
\bar{H}_{S0} & \bar{H}_{SS} & \bar{H}_{SD} \\
\bar{H}_{D0} & \bar{H}_{DS} & \bar{H}_{DD}
\end{pmatrix}
\end{equation}
with $\bar{H}_{SS}$, $\bar{H}_{SD}$, $\bar{H}_{DS}$ and $\bar{H}_{DD}$ are single-single, single-double, double-single and double-double blocks of the matrix. and the $\bar{H}_{S0}$ and $\bar{H}_{D0}$ blocks are zero as they constitute the single and double amplitude residual equation. For (d)rCCD ground state, we note that the single excitations coefficients are absent and the condition $\bar{H}_{S0}=0$ has not been used to determine them. This paves the way for us to consider EOM-CC like formulation for rCCD and drCCD based excited state methods. 

Following the standard derivation of EOM-CC,\cite{shavitt2009} excitation energies limited to single excited states can be obtained as,
\begin{equation}
[\bar{H}_{SS},R_k ]\ket{0}= \omega_{k} R_k \ket{0}
\end{equation}
where $R_k = r_{i}^{a} \{a^{\dagger}{i}\}$ is a single excitation operator,  $\bar{H}_{SS}$ is the effective Hamiltonian for rCCD or drCCD and $\omega_k $ is the excitation energy. The relevant matrix elements for the contraction could be written as
\begin{equation}
[\bar{H}_{SS}{R}]_{i}^{a} = \bar{H}_{e}^{a} r_{i}^{e} -\bar{H}_{m}^{i}r_{m}^{a} + \bar{H}_{ei}^{ma} r_{m}^{e}
\end{equation}
\begin{equation}
[\bar{H}_{SS}{R}]_{ai,bj}= \bar{H}_{e}^{b} - \bar{H}_{m}^{j} + \bar{H}_{bj}^{ia}    
\end{equation}
Substituting for $\bar{H}$ elements we get,
\begin{equation}
[\bar{H}_{SS}{R}]_{ai,bj}= (\epsilon_e - \epsilon_m)\delta_{ae} \delta_{im} + {\bra{ia}}{\ket{bj}} + t_{im}^{ae} {\bra{mj}}{\ket{eb}}
\end{equation}
Since $A_{ai,bj}=(\epsilon_e - \epsilon_m)\delta_{ae} \delta_{im} + {\bra{ia}}{\ket{bj}}$ and $B_{bj,em}={\bra{mj}}{\ket{eb}} $, we obtain 
\begin{equation}
[\bar{H}_{SS}{R}]_{ai,bj} = A_{ai,bj} + B_{bj,em}{t_{ai}^{em}} = A + BT
\label{H_SS}
\end{equation}
As previously shown (d)rCCD residual equation satisfies, 
\begin{equation}
(T ^ {-1} )
\begin{pmatrix}
A & B \\
-B & -A
\end{pmatrix}
\begin{pmatrix}
1 \\
T
\end{pmatrix} =  T^{-1} \begin{pmatrix}
1 \\
T
\end{pmatrix} R
\end{equation}
where $R = A + BT $. Given the eigenvectors and the corresponding eigenvalues of R are X and $ \omega $ respectively, we can write
\begin{equation}
R=X{\omega}X^{-1} 
\end{equation}

Comparison of equations (\ref{RPAeigen}) and (\ref{H_SS}) shows that the EOM-CC formulation of (d)rCCD leads to the corresponding RPA formulation for the excited states. Unlike the canonical derivation of the RPA equations where the ground state is left arbitrary, this approach unambiguously establishes rCCD (drCCD) to be the ground state for RPA (dRPA).  In the context of RPA, this is an important finding. This derivation differs from the approach taken by Berkelbach who has also explored the EOM-CC formulation of rCCD recently and arrived at similar conclusions. The one particle effective Hamiltonian elements, if constructed assuming $\bar{H}= e^{-T}He^{T} $ form for the (d)rCCD approximations, would be  
\begin{equation}
\bar{H}_{m}^{i}=f_{im} \delta_{im} + \dfrac{1}{2} t_{nm}^{fa} {\bra{ni}}{\ket{fa}}
\end{equation}
\begin{equation}
\bar{H}_{e}^{a} = f_{ae} \delta_{ae} - \dfrac{1}{2} t_{mn}^{fa} {\bra{mn}}{\ket{fe}}
\end{equation}
But as we have seen from our discussion, it is not precisely correct to assume that form for (d)rCCD. Instead, the expressions for $\bar{H}$ elements have to be obtained from the corresponding Lagrangian equations. When the $\bar{H}$ elements are strictly r/drCCD, as we have shown that the EOM-CC formulation naturally leads to RPA with the r/drCCD being the proper ground state. This has also been discussed by Berkelbach who states that the equivalence between RPA and EOM-CC approaches occur only if the \textit{one-particle part of the Hamiltonian is \textbf{not} similarity transformed}. In general, $\bar{H}$ for a regular CC calculation is 
\begin{align}
\bar{H}& =  e^{-T}He^{T} \\ \notag
            & =  e^{-T}(F+W)e^{T} \\ \notag
            & =  e^{-T}Fe^{T}  + e^{-T}We^{T}    
\end{align}

\begin{equation}
\bar{H}  = \bar{F} + \bar{W}
\end{equation}
But as we have shown above for rCCD and drCCD, the bare Fock matrix must be used i.e. we use $F$ instead of $\bar{F}$. This would mean
\begin{align}
\bar{H}_{SS} = <S|\bar{H}|S>& = \bra{\Phi_{i}^{a}}\bar{H}\ket{\Phi_{j}^{b}} \\ \notag
& = f_{ij}\delta{ij} - f_{ab}\delta{ij} + \bar{W}_{iajb}
\end{align}
where $\bar{W}_{iajb} = {\bra{ia}}{\ket{jb}} + \Sigma {{t_{jm}^{eb}}{{\bra{mi}}{\ket{eb}}}}$.
We will denote such approximations as EOM(\textbf{Sf}); \textbf{f} underscoring the fact that the Fock matrix elements ($f_{pq}$) are not dressed and \textbf{S} highlighting that diagonalization space is composed of only singly excited determinants as is also the case for RPA eigenvalue problem. If instead, similarity transformed Fock operator ($\bar{F}$) is used, we refer to those approximations as EOM(\textbf{S}). The justification for use of the latter formulation might be on numerical grounds which we intend to assess in this study. 

Following a notation used by Berkelbach,\cite{Berkelbach2018} our parent models in this study are EOM(Sf) rCCD and EOM(Sf) drCCD where the one-particle $\bar{H}$ elements limited to Fock-diagonals  $\bar{H}_{e}^{a} =f_{ae}$ and $\bar{H}_{m}^{i}= f_{mi}$. It has been shown above and in the work of Berkelbach that these EOM models are methodological equivalent of RPA and d-RPA respectively. As mentioned before, we augment these parent models by using dressed one-particle $\bar{H}$ elements, and call the resulting approximations EOM(S)rCCD and EOM(S) drCCD.


\subsection{Doubles correction to EOM(S)-(d)rCCD and EOM(SF)-(d)rCCD models}
The EOM ansatz, $R_k = r_{0} + r_{i}^{a}\{a^{\dagger}i\} $, limited to a single excitation operator and the (d)rCCD ground state leads to the EOM models presented above, and it is obvious that they are \textit{single excitation models}. In contrast, the configuration interaction singles (CIS) is a single excitation model of the HF ground state. There are numerous studies documenting the deficiencies of single excitation models like CIS and RPA (which is of course EOM(Sf)rCCD) predicting excitation spectra. The improvements have been formulated by considering the perturbative inclusion of double excitation effects as in CIS(D) of Head-Gordon and co-workers\cite{headgordon1994} and RPA(D) of Sauer and co-workers.\cite{Sauer1998} Drawing from our experience in formulating the perturbative correction to EOM-CCSD to account for the effect of triple excitations, EOM-CCSD(T)\cite{watts1995, watson2013}, we present a perturbative doubles correction to the EOM (d)rCCD models.  

Let us define the projections $ \ket{P} = \ket{0} + \ket{S} $  and $ \ket{Q} = \ket{D} + \ket{T} + \dots $ where $\ket{S}$, $\ket{D}$, $\ket{T}$  are single, doubly and triply excited configurations. We also note that $\bra{Q}\ket{P} = 0 $. The EOM (d)rCCD equations can be approximately written in terms of the projection operators as,
\begin{equation}
\begin{pmatrix}
\bar{H}_{PP} & \bar{H}_{PQ} \\
\bar{H}_{QP} & \bar{H}_{QQ}
\end{pmatrix}
\begin{pmatrix}
{R_k}_P \\
{R_k}_Q
\end{pmatrix}
= {\omega}
\begin{pmatrix}
{R_k}_P \\
{R_k}_Q
\end{pmatrix}
\end{equation}
We note that except for the $\bar{H}_{PP}$ block, all the other blocks are approximations. As a result, to a good approximation the $\bar{H}$ elements in these blocks are limited to the corresponding bare integrals (those are the lead terms in expressions for $\bar{H}$). Using the Löwdin partitioning technique\cite{Lowdin1963}, this can be rewritten as
\begin{equation}
\bar{H}_{PP}{R_k}_P + \dfrac{\bar{H}_{PQ}\bar{H}_{QP}}{(\omega_{k} - \bar{H}_{QQ})}{{R_k}_P} = \omega_k {R_k}_P
\end{equation}
where ${R_k}_P = \bra{P}\ket{R_k} $. It is clear from the above expression the correction to $\omega_k$ come from $\bar{H}_{PQ}(\omega_{k} - \bar{H}_{QQ})^{(-1)}\bar{H}_{QP}$. Following the steps outlined previously, it can be shown that the lowest order correction terms due to double excitation effects arise from the expansion,
\begin{equation}
\Delta{\omega} = \dfrac{\bar{H}_{PD}\bar{H}_{DP}  }{(\omega_{k} - \bar{H}_{DD})}
\end{equation}
with $\Delta{\omega}$ designating the correction due to the double excitations. This can be rewritten in more convenient form as,
\begin{equation}
\Delta{\omega} = \dfrac{{}{\bra{0}{L_k}\bar{H}\ket{D}} {\bra{D}\bar{H}{R_k}\ket{0}}}{{(\omega_k - \bra{D}\bar{H}\ket{D})}}
\end{equation}
where $L_k$ is the left state corresponding to the right state $R_k$. 
The algebraic expression for $ L_{ij,ab} = \bra{0}{L_k}\bar{H}\ket{D}$ is given by,
\begin{equation}
L_{ij,ab} = P_{-}(ij){l_{ie}}{\bra{ej}}{\ket{ab}} - P_{-}(ab){l_{ma}}{\bra{ij}}{\ket{mb}}
\end{equation}
and similarly for $R_{ab,ij} = \bra{D}\bar{H}{R_k}\ket{0}$,
\begin{equation}
R_{ab,ij}= P_{-}(ij)r_{ei}{\bra{ab}}{\ket{ej}} -P_{-} (ab) {r_{am}} {\bra{mb}}{\ket{ij}}
\end{equation}
Therefore, the doubles correction,
\begin{equation}
\Delta{\omega} = \dfrac{{L_{ij,ab} R_{ab,ij}}}{4 D_{ijab} }    
\end{equation}
where $(\omega_k - \bra{D}\bar{H}\ket{D})$ is $ D_{ijab} = (\omega_k-(\epsilon_a+\epsilon_b-\epsilon_i-\epsilon_j )). $  We assign the labels EOM(Sf, +D) rCCD, EOM(Sf,+D) drCCD, EOM(S,+D) drCCD and EOM(S,+D) drCCD to indicate that the perturbative effects due to double excitation are included. This correction term appears similar to the lead correction term in CIS(D). It appears that the only difference is that in CIS $L_k$ is identical to $R_k$ (CIS being a symmetric theory). The second term in the CIS(D) correction arises from a contribution, 
\begin{equation}
c_{m}^{e} \bar{H}_{ej}^{mb} c_{j}^{b}
\end{equation}
where $ \bar{H}_{ej}^{mb} = \dfrac{1}{2} t_{jn}^{fb} {\bra{mn}}{\ket{ef}}$ and $c_{i}^{a}$ are CIS coefficients. In RPA, this contribution is included to infinite order (i.e. in the RPA matrix) as a term arising from the B matrix. Therefore, in RPA or in EOM(Sf) (d)rCCD, the only perturbative double correction is $\Delta {\omega}$ given above. 

\subsection{Exchange corrections to EOM-CC based on direct-ring CCD ground state}
For a direct-ring CCD based excited state calculation, we remove the exchange piece in $\bar{W}_{iajb}$ and we refer to the method as EOM(S) drCCD and when we keep the exchange piece, we shall call it EOM(S,+X) drCCD. Finally, a third option (not studied in this work) could be to add the exchange piece as a non-iterative correction at the end in the spirit of second order exchange(SOX) and second order screened exchange (SOSEX) correction to the ground state drCCD energy. Such a problem does not arise for ring CCD based EOM calculation (EOM(S) rCCD) where we naturally have the complete $\bar{W}_{iajb}$ with exchange piece intact. We could make the choice to eliminate it completely (EOM(S) rCCD-X) or eliminate it before diagonalization and put it back as a non iterative correction (EOM(S) rCCD(X)) but we have not assessed those corrections for rCCD based EOM model here. 

Our aim in this work is to study all these different approximations and compare them with EOM CCSD results. We intend to find out if any one of these methods have advantages associated with drCCD(dRPA). Our aim will also be to explore how to improve the performance of EOM based on the rCCD model though there are significant issues with the convergence of rCCD ground state calculations. We have found that RPA or the ring-CCD equations are prone to instability and fail to converge in many cases. An instability of the reference determinant in any one of the irreducible representations in the point group symmetry is diagnosed to be the cause of non-convergence. In contrast, the solution to dRPA or direct ring-CCD equations converge in general. As a \textit{workaround}, whenever ring-CCD equations do not converge, we solve the RPA eigenvalue equation for single excitation and de-excitation operators ($X$, $Y$) which are then used to construct the $T_2$ amplitudes and $\bar{H}$ for the EOM-CC problem. Thus, an alternate route to getting to excitation energies is constructed which ends up giving the same excitation energies as RPA. This option only gives a partial solution since unlike the rCCD equations the RPA equations can be solved for each symmetry block separately and those symmetry blocks that suffer from instabilites need not to be considered. Most importantly though, we can procced to obtain the doubles corrections within the our EOM formulation of RPA. This and other considerations in the design of the methods studied in this paper are illustrated through a flowchart in figure \ref{Flowchart_EOMCC_RPA}. In the next section, we focus our attention on numerical performance of these EOM (d)rCCD approximations.

\section{Implementation and Results}\label{Results}

All the methods mentioned are implemented in the development version of software packages \textbf{{ACES II}} \cite{ACESII, ACESpaper}  and \textbf{{Massively Parallel Quantum Chemistry (MPQC)}} \cite{MPQC, MPQCpaper}. To test these methods, we did a simple study of singlet excited states of small closed-shell molecules such as H$_2$O, N$_2$, Ne, CH$_2$ and BH in modified cc-pVDZ basis augmented with diffuse basis functions which were used in an earlier study by Christiansen, Sauer and co-workers.\cite{Christiansen1996b, Sauer1998} 

\subsection{Effect of dressing of Fock matrix elements on RPA excitation energies}

\begin{table}[h!]
\scriptsize
\caption{Effect of dressing of Fock matrix elements, $\Delta_{Fock Dressing}$, on the excitation energies of various small molecules in aug-cc-pVDZ* basis.  }\label{water_excitation_energies}
\begin{tabular}{llllllllll}
\hline
Molecule    & State & FCI   & EOM CCSD & EOM(Sf) rCCD & EOM(S) rCCD & $\Delta_{Fock Dressing}^{EOMrCCD}$ & EOM(Sf) drCCD & EOM(S) drCCD & $\Delta_{Fock Dressing}^{EOMdrCCD}$ \\
    &  &    &  & (RPA) &  &  & (dRPA) & &  \\

\hline
 H$_{2}$O & $1 ^{1}{B_1}$ & 7.45  & 7.38  & 8.63  & 10.33 & 1.70 & 14.93 & 15.48 & 0.56 \\
 & $1 ^{1}{A_2}$ & 9.21  & 9.12  & 10.32 & 12.05 & 1.74 & 15.47 & 16.03 & 0.55 \\
 & $2 ^{1}{A_1}$ & 9.87  & 9.81  & 10.95 & 12.62 & 1.67 & 17.01 & 17.58 & 0.57 \\
 & $1 ^{1}{B_2}$ & 11.61 & 11.52 & 12.61 & 14.29 & 1.69 & 17.54 & 18.11 & 0.57 \\
\hline
N$_{2}$ & $^{1}{\Pi}_g$ & 9.58 & 9.66 & 9.82 & 12.74 & 2.92 & 23.43 & 24.38 & 0.95 \\
 & $^{1}{\Sigma}_u$ & 10.33 & 10.47 & 8.09 & 12.01 & 3.92 & 21.42 & 23.89 & 2.47 \\
 & $^{1}{\Delta}_u$ & 10.72 & 10.90 & 8.96 & 10.02 & 1.06 & 22.74 & 23.89 & 1.15 \\
\hline
Ne & $^{1}P$ & 16.40 & 16.16 & 18.09 & 19.18 & 1.09 & 25.51 & 26.08 & 0.57 \\
 & $^{1}D$ & 18.21 & 17.96 & 19.94 & 21.03 & 1.09 & 24.99 & 25.55 & 0.56 \\
 & $^{1}S$ & 18.26 & 18.01 & 19.98 & 21.06 & 1.08 & 25.03 & 25.59 & 0.56 \\
 \hline
CH$_{2}$ & $1 ^{1}B_{2}$ & 1.79 & 1.78 & 1.64 & 2.37 & 0.73 & 13.74 & 14.26 & 0.52 \\
 & $1 ^{1}A_{2}$ & 5.85 & 5.86 & 6.03 & 6.88 & 0.85 & 17.75 & 18.29 & 0.54 \\
 & $3 ^{1}A_{1}$ & 6.51 & 6.51 & 6.94 & 7.68 & 0.74 & 11.51 & 11.92 & 0.41 \\
 & $1 ^{1}B_{1}$ & 7.70 & 7.71 & 8.10 & 8.83 & 0.73 & 12.37 & 12.78 & 0.41 \\
 & $4 ^{1}A_{1}$ & 8.48 & 8.46 & 8.84 & 9.55 & 0.71 & 12.83 & 13.25 & 0.41 \\
 \hline
BH & $A ^{1}{\Pi}^{+}$ & 2.94 & 2.96 & 2.85 & 3.20 & 0.35 & 9.77 & 10.12 & 0.35 \\
 & $B ^{1}{\Sigma}^{+}$ & 6.38 & 6.42 & 6.37 & 6.72 & 0.35 & 9.67 & 10.01 & 0.35 \\
 & $D ^{1}\Pi$ & 7.47 & 7.50 & 7.36 & 7.70 & 0.34 & 10.80 & 11.16 & 0.36 \\
 & $E ^{1}{\Sigma}^{+}$ & 7.56 & 7.39 & 7.38 & 7.72 & 0.34 & 9.80 & 10.15 & 0.35 \\
 & $G ^{1}{\Pi}$ & 8.24 & 8.28 & 8.11 & 8.44 & 0.33 & 11.68 & 12.03 & 0.35 \\
 \hline
\end{tabular}
\end{table}

Previous studies have documented the accuracy of excitation energies from RPA method and found it to be slightly better than configuration interaction singles (CIS) method which are in error of $1$ to $1.5$ eV  for singly excited states. Our test results for small molecules (See results for water in Table \ref{water_excitation_energies}) reiterate the previous findings. As RPA excitation energies are equivalent to those of  EOM(Sf)-rCCD methods which uses bare Fock matrix elements, our aim is to quantify the effect of \textit{similarity transformation of Fock operators} or \textit{dressing of Fock matrix elements} on the excitation energies. The results below indicate the magnitude of $\Delta_{Fock Dressing}^{EOM rCCD}$, to be significantly large ($\approx$ $1.70$ eV) and is indicated to worsen the estimate of excitation energy further. This result seems to be in agreement with recent work by Lange and Berkelbach who investigated the effect of categories of diagrams present in EOM-CCSD but absent in RPA and concluded that \textit{high-quality vertex corrections (from EOM-CCSD) to the polarizability do not improve the ionization potentials of small molecules within the GW approximation}. \cite{Berkelbach2019, Berkelbach2018b} 

Another observation is about the direct RPA (dRPA) excitation energies which are not adequately reported for molecular systems. EOM(Sf) drCCD (dRPA) estimates are $\approx$ $5-6$ eV worse than EOM(Sf) rCCD. Interestingly, the effect of dressing of Fock matrix elements through similarity transformation is considerably less, $\Delta_{Fock Dressing}^{EOM drCCD}$ $\approx$ $0.6$ eV, in this case. In the next subsection, we seek improvement in dRPA/EOM(SF) drCCD method  by adding exchange corrections to $\bar{H}$ elements for drCCD.

\subsection{Addition of exchange corrections to EOM based on direct ring CCD ground state calculations}
\begin{table}[ht!]
\scriptsize
\caption{Effect of adding exchange corrections to EOM models based on direct ring CCD ground state calculation }
\begin{tabular}{lllllllllll}
\hline
Molecule & State & EOM CCSD & CIS   & EOM(Sf) drCCD & EOM(Sf,+X) drCCD & $ \Delta_{Exchange}^{SF} $ & EOM(S) drCCD & EOM(S,+X) drCCD & $ \Delta_{Exchange}^{S} $ \\
 &  &  &   & (dRPA) &  &  &  &  & \\
\hline
H$_2$O & $1 ^{1}{B_1}$ & 7.38 & 8.67 & 14.93 & 8.66 & 6.26 & 15.48 & 9.29 & 6.19 \\
 &$1 ^{1}{A_2}$ & 9.12 & 10.36 & 15.47 & 10.36 & 5.11 & 16.03 & 11.00 & 5.03 \\
 &  $2 ^{1}{A_1}$ & 9.81 & 10.98 & 17.01 & 10.98 & 6.03 & 17.58 & 11.61 & 5.97 \\
 & $1 ^{1}{B_2}$ & 11.52 & 12.64 & 17.54 & 12.64 & 4.90 & 18.11 & 13.27 & 4.83 \\
 \hline
N$_2$ & $^{1}{\Pi}_g$ & 9.66 & 10.07 & 23.43 & 10.04 & 13.39 & 24.38 & 11.06 & 13.32 \\
 & $^{1}{\Sigma}_u$ & 10.47 & 8.65 & 21.42 & 8.65 & 12.77 & 23.89 & 9.86 & 14.03 \\
 & $^{1}{\Delta}_u$ & 10.9 & 9.23 & 22.74 & 9.27 & 13.46 & 23.89 & 10.48 & 13.40 \\
 \hline
Ne & $^{1}P$ & 16.16 & 18.09 & 25.51 & 18.09 & 7.43 & 26.08 & 18.66 & 7.42 \\
 & $^{1}D$ & 17.96 & 19.94 & 24.99 & 19.95 & 5.04 & 25.55 & 20.51 & 5.04 \\
 & $^{1}S$ & 18.01 & 19.98 & 25.03 & 19.98 & 5.05 & 25.59 & 20.55 & 5.04 \\
 \hline
CH$_2$ & $1 ^{1}B_{2}$ & 1.78 & 1.64 & 13.74 & 1.60 & 12.15 & 14.26 & 2.13 & 12.13 \\
 & $1 ^{1}A_{2}$ & 5.86 & 6.07 & 17.75 & 6.06 & 11.69 & 18.29 & 6.61 & 11.68 \\
 & $3 ^{1}A_{1}$ & 6.51 & 6.95 & 11.51 & 6.95 & 4.56 & 11.92 & 7.38 & 4.54 \\
 & $1 ^{1}B_{1}$ & 7.71 & 8.11 & 12.37 & 8.12 & 4.25 & 12.78 & 8.55 & 4.23 \\
 & $4 ^{1}A_{1}$ & 8.46 & 8.86 & 12.83 & 8.86 & 3.97 & 13.25 & 9.28 & 3.97 \\
 \hline
BH & $A ^{1}{\Pi}^{+}$ & 2.96 & 2.84 & 9.77 & 2.80 & 6.97 & 10.12 & 2.19 & 7.93 \\
 & $B ^{1}{\Sigma}^{+}$ & 6.42 & 6.37 & 9.67 & 6.37 & 3.30 & 10.01 & 5.74 & 4.28 \\
 & $D ^{1}\Pi$ & 7.5 & 7.36 & 10.80 & 7.37 & 3.43 & 11.16 & 6.93 & 4.22 \\
 & $E ^{1}{\Sigma}^{+}$ & 7.39 & 7.39 & 9.80 & 7.39 & 2.41 & 10.15 & 6.93 & 3.21 \\
 & $G ^{1}{\Pi}$ & 8.28 & 8.12 & 11.68 & 8.11 & 3.57 & 12.03 & 7.71 & 4.32 \\
 \hline
\end{tabular}\label{ExchangeCorrections}
\end{table}

In contrast to the ground state correlation energies for which drCCD method seems to be provide qualitatively correct estimates, we find poor results for its excited state analog. As there has been a lot of recent work on how to improve drCCD correlation energies by adding exchange corrections perturbatively,\cite{Gruneis2009,Hummel2019} we try similar corrections to some $\bar{H}$ matrix elements for drCCD ground state. To be abundantly clear, we do not add these corrections to the ground-state amplitude equations or ground state energy, but add them post ground state calculations while constructing $\bar{H}$ before the diagonalization step. 

As we see in Table {\ref{ExchangeCorrections}}, addition of exchange terms in $\bar{W}_{iajb}^{drCCD}$ leads to significant recovery in accuracy of these methods. Such exchange corrections ($ \Delta_{Exchange}^{S/SF} $) improve the estimates and we find that the EOM(Sf,+X)-drCCD values tends to be very close to EOM(Sf)-rCCD. This is not expected as ground state amplitudes and energies obtained from drCCD and rCCD methods tend to be different. The differences in the corresponding $T$ amplitudes would be more severe when away from equilibrium geometries or in static correlation dominated systems. It will be interesting to check the performance of EOM-drCCD models in such cases. We also find the exchange correction for EOM(Sf) and EOM(S) variants, $\Delta_{Exchange}^{SF} \approx \Delta_{Exchange}^{S}$, to be similar. The performance of EOM(S,+X) drCCD method, which still behaves worse than CIS, is puzzling. In the next subsection, we go beyond single excitation space (S) and analyze the contribution of double excitations (D) in a perturbative manner.

\subsection{Perturbative effect of double excitations on the energies of singly excited states}

\begin{table}[h!]
\scriptsize
\caption{Perturbative estimate of the effect of doubles excitations added to EOM models (with exchange correction) based on direct ring CCD ground state calculation }

\begin{tabular}{lllllllllll}
\hline
Molecule & State &  EOM CCSD & CIS   & EOM(Sf,+X)  & EOM(Sf,+X,+D)  & $ \Delta_{Doubles} ^ {SF, +X, drCCD}$ & EOM(S,+X)  & EOM(S,+X,+D)  & $ \Delta_{Doubles}^{S,+X, drCCD} $ \\
 &  &  &   &  drCCD & drCCD &  &drCCD &  drCCD &   \\
\hline
H$_{2}$O & $1  ^{1}B_{1}$ & 7.38 & 8.67 & 8.66 & 5.19 & -3.47 & 9.29 & 5.75 & -3.54 \\
 & $1  ^{1}A_{2}$ & 9.12 & 10.36 & 10.36 & 6.88 & -3.48 & 11.00 & 7.43 & -3.57 \\
 & $2 ^{1}A_{1}$ & 9.81 & 10.98 & 10.98 & 7.62 & -3.36 & 11.61 & 8.18 & -3.43 \\
 & $1 ^{1}B_{2}$ & 11.52 & 12.64 & 12.64 & 9.25 & -3.39 & 13.27 & 9.80 & -3.47 \\
  \hline
N$_{2}$ & $^{1}{\Pi}_g$ & 9.66 & 10.07 & 10.04 & 6.85 & -3.19 & 11.06 & 7.81 & -3.25 \\
 & $^{1}{\Sigma}_u$ & 10.47 & 8.65 & 8.65 & 6.90 & -1.74 & 9.86 & 8.10 & -1.76 \\
 & $^{1}{\Delta}_u$ & 10.9 & 9.23 & 9.27 & 7.42 & -1.85 & 10.48 & 8.61 & -1.87 \\
  \hline
Ne & $^{1}P$ & 16.16 & 18.09 & 18.09 & 14.00 & -4.09 & 18.66 & 14.54 & -4.12 \\
 & $^{1}D$ & 17.96 & 19.94 & 19.95 & 15.67 & -4.27 & 20.51 & 16.21 & -4.31 \\
 & $^{1}S$ & 18.01 & 19.98 & 19.98 & 15.72 & -4.26 & 20.55 & 16.25 & -4.29 \\
  \hline
CH$_{2}$ & $1 ^{1}B_{2}$ & 1.78 & 1.64 & 1.60 & 0.29 & -1.30 & 2.13 & 0.82 & -1.31 \\
 & $1 ^{1}A_{2}$ & 5.86 & 6.07 & 6.06 & 4.48 & -1.58 & 6.61 & 5.02 & -1.60 \\
 & $3 ^{1}A_{1}$ & 6.51 & 6.95 & 6.95 & 5.29 & -1.66 & 7.38 & 5.70 & -1.68 \\
 & $1 ^{1}B_{1}$ & 7.71 & 8.11 & 8.12 & 6.47 & -1.65 & 8.55 & 6.87 & -1.67 \\
 & $4 ^{1}A_{1}$ & 8.46 & 8.86 & 8.86 & 7.27 & -1.59 & 9.28 & 7.67 & -1.61 \\
 \hline
BH & $A ^{1}{\Pi}^{+}$ & 2.96 & 2.84 & 2.80 & 1.76 & -1.03 & 3.23 & 2.19 & -1.04 \\
 & $B ^{1}{\Sigma}^{+}$ & 6.42 & 6.37 & 6.37 & 5.39 & -0.98 & 6.73 & 5.74 & -0.99 \\
 & $D ^{1}\Pi$ & 7.5 & 7.36 & 7.37 & 6.59 & -0.78 & 7.72 & 6.93 & -0.79 \\
 & $E ^{1}{\Sigma}^{+}$ & 7.39 & 7.39 & 7.39 & 6.51 & -0.88 & 7.72 & 6.93 & -0.79 \\
 & $G ^{1}{\Pi}$ & 8.28 & 8.12 & 8.11 & 7.37 & -0.74 & 8.46 & 7.71 & -0.75 \\
\hline

\end{tabular}\label{DoublesCorrectionstoEOMdrCCD}
\end{table}

A computationally cheap way to include the effect of double excitations, without increasing the diagonalization space for the effective Hamiltonian $\bar{H}$, is to add a perturbative correction. A variation of a non-iterative doubles correction proposed by Head-Gordon \textit{et al.} for the CIS method,\cite{headgordon1994} termed the (D) correction, is considered here. We shall consider the various ways of adding this correction to the set of EOM-(d)rCCD methods. 

\begin{table}[h!]
\scriptsize
\caption{Perturbative estimate of the effect of doubles excitations added to EOM models (with exchange correction) based on ring CCD ground state calculation }
\begin{tabular}{llllllllllll}
\hline
Water & State & EOM CCSD & CIS   & EOM(Sf) rCCD & EOM(Sf, +D) rCCD & $ \Delta_{Doubles} ^ {SF, rCCD}$ & EOM(S) rCCD & EOM(S, +D) rCCD & $ \Delta_{Doubles} ^ {S, rCCD}$ \\
\\
\hline
H$_{2}$O & $1 ^{1}B_{1}$ & 7.38 & 8.67 & 8.63 & 5.17 & -3.46 & 10.33 & 6.67 & -3.65 \\
 & $1 ^{1}A_{2}$ & 9.12 & 10.36 & 10.32 & 6.86 & -3.45 & 12.05 & 8.33 & -3.72 \\
 & $2 ^{1}A_{1}$ & 9.81 & 10.98 & 10.95 & 7.60 & -3.36 & 12.62 & 9.09 & -3.53 \\
 & $1 ^{1}B_{2}$ & 11.52 & 12.64 & 12.61 & 9.23 & -3.38 & 14.29 & 10.69 & -3.60 \\
 \hline
N$_{2}$ & $^{1}{\Pi}_g$ & 9.66 & 10.07 & 9.82 & 6.77 & -3.05 & 12.74 & 9.53 & -3.22 \\
 & $^{1}{\Sigma}_u$ & 10.47 & 8.65 & 8.09 & 6.74 & -1.35 & 12.01 & 10.84 & -1.17 \\
 & $^{1}{\Delta}_u$ & 10.90 & 9.23 & 8.96 & 7.22 & -1.74 & 12.86 & 11.06 & -1.80 \\
 \hline
Ne & $^{1}P$ & 16.16 & 18.09 & 18.09 & 14.01 & -4.08 & 19.18 & 15.04 & -4.13 \\
 & $^{1}D$ & 17.96 & 19.94 & 19.94 & 15.67 & -4.27 & 21.03 & 16.69 & -4.34 \\
 & $^{1}S$ & 18.01 & 19.98 & 19.98 & 15.80 & -4.18 & 21.06 & 16.85 & -4.20 \\
 \hline
CH$_{2}$ & $1 ^{1}B_{2}$ & 1.78 & 1.64 & 1.64 & 0.38 & -1.26 & 2.37 & 1.09 & -1.28 \\
 & $1 ^{1}A_{2}$& 5.86 & 6.07 & 6.03 & 4.49 & -1.54 & 6.88 & 5.31 & -1.57 \\
 & $3 ^{1}A_{1}$ & 6.51 & 6.95 & 6.94 & 5.30 & -1.64 & 7.68 & 6.00 & -1.68 \\
 & $1 ^{1}B_{1}$ & 7.71 & 8.11 & 8.10 & 6.48 & -1.62 & 8.83 & 7.16 & -1.67 \\
 & $4 ^{1}A_{1}$ & 8.46 & 8.86 & 8.84 & 7.27 & -1.57 & 9.55 & 7.94 & -1.61 \\
 \hline
BH & $A ^{1}{\Pi}^{+}$ & 2.96 & 2.84 & 2.85 & 1.87 & -0.98 & 3.20 & 2.21 & -0.99 \\
 & $B ^{1}{\Sigma}^{+}$ & 6.42 & 6.37 & 6.37 & 5.40 & -0.97 & 6.72 & 5.75 & -0.97 \\
 & $D ^{1}\Pi$ & 7.50 & 7.36 & 7.36 & 6.65 & -0.71 & 7.70 & 6.97 & -0.73 \\
 & $E ^{1}{\Sigma}^{+}$ & 7.39 & 7.39 & 7.38 & 6.51 & -0.87 & 7.72 & 6.84 & -0.88 \\
 & $G ^{1}{\Pi}$ & 8.28 & 8.12 & 8.11 & 7.42 & -0.69 & 8.44 & 7.74 & -0.70 \\
\hline
\end{tabular}\label{DoublesCorrectionstoEOMrCCD}
\end{table}

We find that the doubles correction $\Delta_{Doubles}^{S/SF,+X, drCCD}$ are very large and in opposite direction ($\approx$ $ -0.8$ to $- 4.5 $ eV) as shown in Table \ref{DoublesCorrectionstoEOMdrCCD}, and when added to EOM drCCD energies underestimate the FCI values by $\approx$ $2 - 2.5 $ eV. The (D) correction for $ \Delta_{Doubles} ^ {S/SF, rCCD}$, is of similar magnitude and the effect is similar (Table \ref{DoublesCorrectionstoEOMrCCD}). The most complete of these methods in theoretically, EOM(S,+D) rCCD, has deviations ${0.7-0.9}$ eV.

\subsection{Analysis}
\begin{figure}[ht!bp]
\centering
\includegraphics[scale=0.9]{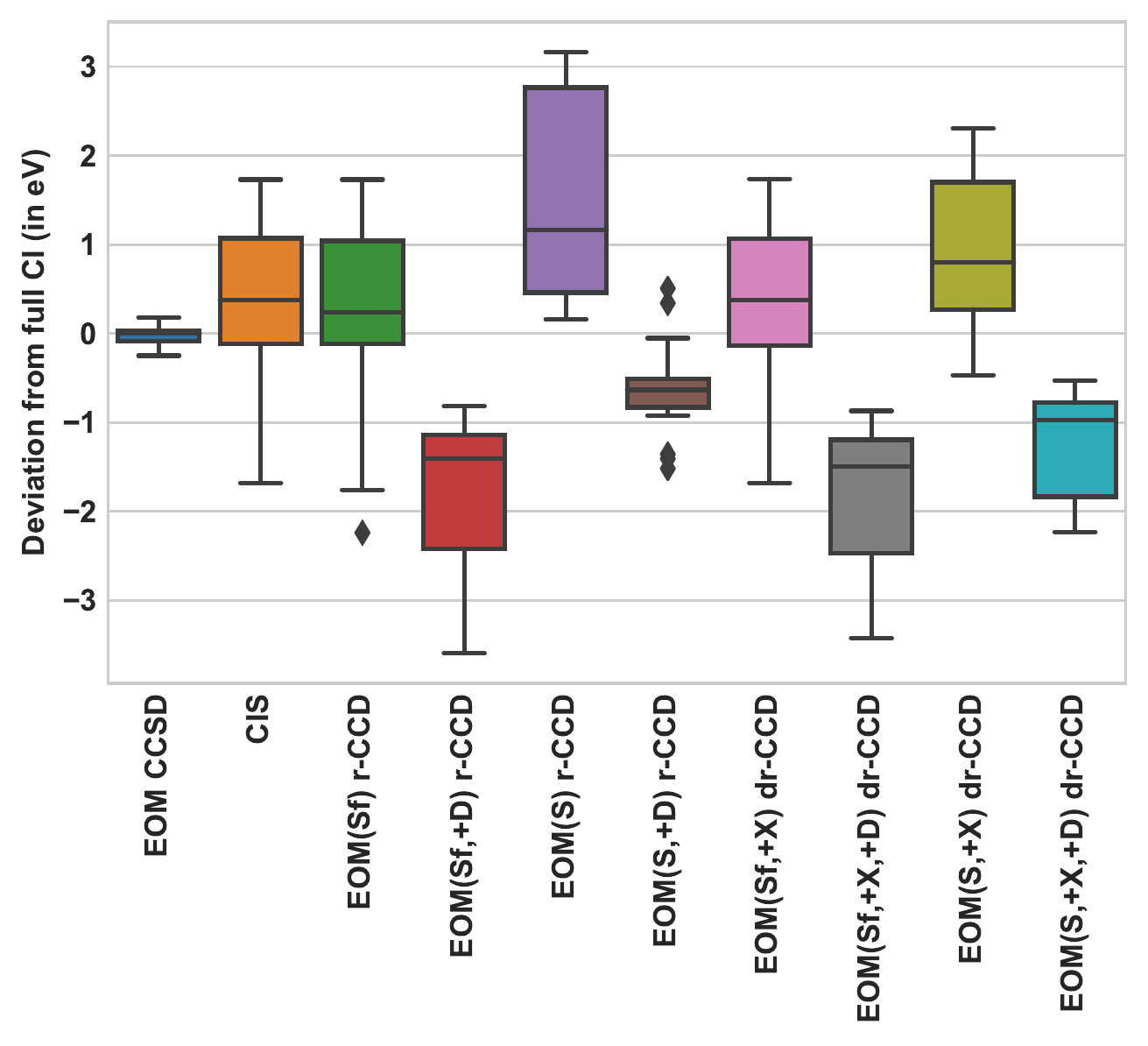}
\caption{Deviations from full CI (in eV) for EOM-CC approximations introduced in the paper} \label{BoxPlot_ErrorSpread}
\end{figure}

We plot the errors in the EOM-CC approximations introduced as compared to the full CI approaches in Fig \ref{BoxPlot_ErrorSpread}. We find EOM(S,+D)-rCCD approach to be amongst the most accurate of the new approximations and an improvement over the CIS method, but significantly less accurate than the EOM-CCSD approach. The error range of various methods is a mark of their consistency; EOM-CCSD shows the least spread followed by EOM(S,+D)-rCCD. Table \ref{Analyze_Stats}  affirms this and points to the the consistency that results by systematic corrections to the parent RPA (EOM(Sf)-rCCD) method.

\begin{table}[h!]
\scriptsize
\caption{Statistical analysis for the studied benchmark set lists MAE(mean absolute error), MSE(mean signed error) and  MAX(maximum error) 
with respect to the FCI values.}
\begin{tabular}{llllllllll}
\hline
\hline
 & EOM & CIS & EOM(Sf)& EOM(Sf, +D) & EOM(S) & EOM(S, +D) & EOM(Sf,+X,+D) & EOM(S,+X)  & EOM(S,+X,+D) \\
  & CCSD &  & rCCD & rCCD & rCCD & rCCD & drCCD &  drCCD & drCCD \\
 \hline
MAE & 0.09 & 0.77 & 0.79 & 1.89 & 1.63 & 0.72 & 1.88 & 1.05 & 1.31 \\
MSE & -0.03 & 0.39 & 0.32 & -1.89 & 1.63 & -0.64 & -1.88 & 0.98 & -1.31 \\
MAX & 0.18 & 1.73 & 1.73 & -0.82 & 3.16 & 0.51 & -0.87 & 2.30 & -0.53 \\
\hline

\end{tabular}
\label{Analyze_Stats}
\end{table}

\subsection{Dependence on reference: performance of EOM (d)-rCCD based on Kohn-Sham determinant}

As RPA is often applied in conjunction with Kohn-Sham DFT (KS- DFT), it is worthwhile to check if the EOM models considered would yield better results when based on KS-DFT single determinants rather than the HF determinant. With both PBE and B3LYP exchange-correlation functionals, we find similar change in EOM-CCSD and CIS making EOM-CCSD results closer to FCI while worsening the CIS estimates. None of the EOM drCCD or the EOM rCCD models improve in accuracy in comparison to previous results. This is surprising and needs to be understood better. The results for water molecule are shown in Table \ref{KS_reference_PBE} and \ref{KS_reference_B3LYP} for EOM-CC based on KS-DFT reference using PBE and B3LYP functional and similar trend is seen for other molecules in the small set studied here. 

\begin{table}[ht!bp]
\caption{KS-DFT reference (PBE) for CIS and EOM-CC models for excited states of water molecule. }
\scriptsize
\begin{tabular}{lllllllllllll}
\hline 
State & FCI & EOM  & CIS & EOM(Sf,+X)  & EOM(S,+X) & EOM(Sf) & EOM(Sf, +D) & EOM(S) & EOM(S, +D)  \\ 
 &  & CCSD &  & drCCD & drCCD & rCCD & rCCD & rCCD & rCCD \\ 
\hline
1  $^{1}B_{1}$ & 7.45 & 7.41 & 8.76 & 8.75 & 9.42 & 11.85 & 8.88 & 10.27 & 6.40 \\
1  $^{1}A_{2}$ & 9.21 & 9.16 & 10.45 & 10.45 & 11.13 & 13.16 & 9.74 & 12.00 & 8.01 \\
2  $^{1}A_{1}$ & 9.87 & 9.84 & 11.11 & 11.10 & 11.78 & 11.07 & 7.53 & 12.61 & 9.86 \\
1  $^{1}B_{2}$ & 11.61 & 11.57 & 12.78 & 12.77 & 13.46 & 13.26 & 10.60 & 14.35 & 10.49 \\
\hline
    
\end{tabular}\label{KS_reference_PBE}
\end{table}

\begin{table}[ht!bp]
\caption{KS-DFT reference (B3LYP) for CIS and EOM-CC models for excited states of water molecule.}
\scriptsize
\begin{tabular}{lllllllllllll}
\hline 
State & FCI & EOM  & CIS & EOM(Sf,+X)  & EOM(S,+X) & EOM(Sf) & EOM(Sf, +D) & EOM(S) & EOM(S, +D)  \\ 
 &  & CCSD &  & drCCD & drCCD & rCCD & rCCD & rCCD & rCCD \\ 
\hline
1  $^{1}B_{1}$ & 7.45 & 7.40 & 8.68 & 8.68 & 9.34 & 11.80 & 8.82 & 10.16 & 6.29 \\
1  $^{1}A_{2}$ & 9.21 & 9.15 & 10.37 & 10.38 & 11.05 & 13.08 & 9.63 & 11.88 & 7.90 \\
2  $^{1}A_{1}$ & 9.87 & 9.83 & 11.02 & 11.01 & 11.68 & 10.99 & 7.45 & 12.55 & 8.81 \\
1  $^{1}B_{2}$ & 11.61 & 11.56 & 12.69 & 12.69 & 13.36 & 13.20 & 10.55 & 14.22 & 10.37 \\ 
 \hline    
\end{tabular}\label{KS_reference_B3LYP}
\end{table}

\subsection{Singlet-Triplet gaps in Ethylene: A challenge for RPA methods}
Though the focus of this work is the study of singlet excitation energies, RPA is known to severely underestimate triplet excitation energies referred to as \textit{triplet instability}. As a consequence, the singlet-triplet (S-T) gaps for molecular systems are also not accurately estimated. An example is the ethylene molecule for which RPA methods have been studied in the past,\cite{Mckoy1971, Zimmerman2017}. We use the S-T gap in this model problem to test our perturbative corrections. The geometry for ethylene were taken from a previous study.\cite{sauer2009}

\begin{table}
\caption{Singlet-Triplet ($B_{3u}$) gap in Ethylene. Vertical excitation energies was calculated in Ahlrichs-TZVP (valence triple zeta) basis.\cite{Ahlrichs1994} As ring CCD ground state calculations did not converge, EOM-rCCD results are not listed.}
\label{S-T gap}
\begin{tabular}{ll}
\hline
 & S-T gap in Ethylene (eV) \\
\hline
Experiment\cite{Zimmerman2017} & 4.3 - 4.6 \\
EOM-CCSD & 4.08 \\
EOM(Sf) drCCD & 2.44 \\
EOM(Sf, +D) drCCD & 2.38 \\
EOM(Sf,+X) drCCD & 4.07 \\
EOM(Sf,+X,+D) drCCD & 4.06 \\
EOM(S) drCCD & 2.38 \\
EOM(S,+D) drCCD & 2.30 \\
EOM(S,+X) drCCD & 4.00 \\
EOM(S,+X,+D) drCCD & 4.00 \\
EOM(Sf) rCCD & Not Converged \\
\hline
\end{tabular}%
\end{table}

Our results show significant improvement to direct RPA estimates ($2.4 eV $) for the S-T gap as addition of exchange corrections and perturbative effects of double excitations lead to values comparable to EOM-CCSD (See Table \ref{S-T gap}). Ring CCD ground state calculations did not converge in this basis and the alternate route through RPA calculations were also not accessible as the instability was found in the same irreducible representation of the symmetry group in which the singlet and triplet states in question lie. A more detailed study focused on singlet-triplet gaps of a large set of molecular systems would be done in future.

\section{Summary}
In summary, we have presented an unified outlook of CC and random phase approximation highlighting their connections and differences. We have assessed the performance of random phase approximation(RPA) and associated methods for the excitation energies of singlet excited states. We have built on the connection of RPA with EOM-CCD to extend the theory to incorporate exchange corrections to direct RPA methods, perturbative effect of double excitations and reference determinant sensitivity. In particular, EOM(S,+D)-rCCD method improves considerably on RPA excitation energies. The current results lay down a baseline for introduction of more corrections as search for a low cost alternative to EOM-CCSD method continues.  In future work, it might be worthwhile to limit the space of excitations to \textit{ph} and \textit{hp} excitations but have the ground state well correlated by CC methods such as CCSD and check the effect on excitation energies. Associated problem is the assessment of the importance of single excitations in RPA (or ring CCD) and how that might affect the excitation energy estimates. A more rigorous test on benchmark sets\cite{sauer2009, Loos2018, Loos2020} for the family of methods introduced would be done in future.


\section{Acknowledgments}

VR thanks Prof. Ed Valeev, his postdoctoral advisor at Virginia Tech, for encouragement to pursue the project and support through U.S. National Science Foundation grants
(Award Nos. 1550456 and 1800348). This work was supported by the United States Army Research Office (ARO Grant No. W911NF-16-1-0260). 

\section{Data Availability}
The relevant data generated during this study is available within the article.  

\section{Appendix}\label{Appendix}

\subsection{Ground state correlation energy through analogous coupled cluster and RPA methods}

The lack of antisymmetry in double excitation amplitudes ($T_2$; $t_{ij}^{ab} \neq -t_{ji}^{ab}$ and  $t_{ij}^{ab} \neq -t_{ij}^{ba} $) in ring CCD has several implications. As in CCD, the spin-orbital form of the rCCD equations has three unique sets of amplitudes, $t_{IJ}^{AB}$, $t_{ij}^{ab}$ and $t_{Ij}^{Ab}$, where uppercase and lowercase letters designate $\alpha$ and $\beta$ spin-orbitals respectively. In open-shell implementations based on unrestricted (UHF) or restricted open-shell HF (ROHF) reference, we solve three set of residual equations corresponding to these three unique set of amplitudes. While for spin-adapted RHF, there are two set of amplitudes: the triplet spin-adapted amplitudes and the singlet spin-adapted amplitudes, and the two corresponding amplitude equations. In contrast, when the amplitudes are antisymmetric, there is only one set of spin-adapted RHF amplitudes (and one corresponding amplitude equation). The UHF and the spin-adapted RHF implementations give identical energies for closed shell systems (assuming that the reference RHF and UHF states too are identical). This is due to $t_{IJ}^{AB} = t_{Ij}^{Ab}- t_{iJ}^{bA}$ (as $t_{Ij}^{Ab} =t_{bA}^{iJ}$ is the only unique amplitude) which no longer holds when the amplitudes are not antisymmetric. This can be seen by comparing the algebraic expressions for the CCD and rCCD energies. The spin-orbital expression for CCD correlation energy is, 
\begin{equation}
E_{CCD}=\dfrac{1}{4} Tr(\bar{B}\bar{T})=\dfrac{1}{2}Tr(B\bar{T})
\end{equation}
which leads to the spin-integrated UHF CCD (UCCD) energy,
\begin{equation}
E_{UCCD} = \dfrac{1}{4} \bar{t}_{IJ}^{AB}{\bra{IJ}}{\ket{AB}} +  \dfrac{1}{4} \bar{t}_{ij}^{ab}{\bra{ij}}{\ket{ab}} + t_{Ij}^{Ab}{\bra{Ij}}{\ket{Ab}}
\end{equation}
while the RHF CCD (RCCD) energy is,
\begin{equation}
E_{RCCD} = \dfrac{1}{2} \bar{t}_{IJ}^{AB} {\bra{IJ}}{\ket{AB}} + t_{Ij}^{Ab} {\bra{Ij}}{\ket{Ab}}
\end{equation}
as ${\bra{IJ}}{\ket{AB}} = {\bra{ij}}{\ket{ab}}$ and $t_{IJ}^{AB} = t_{ij}^{ab}$. The spin-adapted form is  
\begin{equation}
E_{RCCD} = t_{Ij}^{Ab} {(2{\bra{Ij}}{\ket{Ab}} -    {\bra{Ij}}{\ket{bA}})}
\end{equation}
The overbar explicitly identifies the antisymmetric quantities. We can immediately see that for the closed-shell molecules UHF and RHF CCD correlation energies are identical when the underlying reference states are also identical. This does not hold for rCCD. Unlike the CCD energy, the spin-orbital energy,
\begin{equation}
E_{rCCD}= \dfrac{1}{4} Tr(\bar{B}T)    
\end{equation}
can only be expressed in terms of the antisymmetrized integrals $\bar{B}$ since $T$ is no longer antisymmetric. There are two possible UHF and the corresponding spin-adapted RHF formulations. We could neglect the spin-flip excitations\cite{Hehn2013, Hehn2015} and the corresponding UHF or ROHF reference based correlation energy in spin-integrated form is 
\begin{equation}
E_{UrCCD} = \dfrac{1}{4} \bar{t}_{IJ}^{AB}{\bra{IJ}}{\ket{AB}} +  \dfrac{1}{4} \bar{t}_{ij}^{ab}{\bra{ij}}{\ket{ab}} + \dfrac{1}{2}t_{Ij}^{Ab}{\bra{Ij}}{\ket{Ab}}
\end{equation}
while the corresponding RHF spin-adapted energy is,
\begin{equation}
E_{RrCCD} = \dfrac{1}{4} ({^{3}}{B}{^{3}}{T} + {^{1}}{B}{^{1}}{T})
\end{equation}
where singlet and triplet spin-adapted integrals, ${^{1}}{B}$ and ${^{3}}{B}$ are given by ${^{1}}{B} = {2}{\bra{Ij}}{\ket{Ab}} - {\bra{Ij}}{\ket{bA}}$ and ${^{3}}{B} = - {\bra{Ij}}{\ket{bA}}$ respectively. Similarly, the singlet and triplet spin-adapted rCCD amplitudes, ${^{1}}{T}$ and ${^{3}}{T}$ are given by $t_{Ij}^{AB} + t_{Ij}^{Ab}$ and $t_{Ij}^{AB} - t_{Ij}^{Ab}$ respectively. We postpone the discussion of spin-flip rCCD until we present ground state correlation energy expressions for RPA. This approach is more transparent since it is very easy to see the origin of the spin-flip excitations in the context of RPA and then to import those ideas to rCCD. In the case of drCCD, all the two-electron integrals are Coulomb only and symmetric. Therefore the spin-orbital expression for the energy is,
\begin{equation}
E_{drCCD}=\dfrac{1}{2} Tr(BT)
\end{equation}
Here the factor is $\dfrac{1}{2} $ instead of $\dfrac{1}{4}$ because B is symmetric. The spin-integrated UHF/ROHF energy is,
\begin{equation}
E_{UdrCCD} = \dfrac{1}{2} t_{IJ}^{AB}{\bra{IJ}}{\ket{AB}}   + \dfrac{1}{2} t_{ij}^{ab} {\bra{ij}}{\ket{ab}} + t_{Ij}^{Ab} {\bra{Ij}}{\ket{Ab}} 
\end{equation}
The corresponding spin-adapted RHF energy is
\begin{equation}
E_{UdrCCD} = {2} {t_{Ij}^{Ab}}{\bra{Ij}}{\ket{Ab}}
\end{equation}

All the above expressions are derived from the (d)rCCD route. Now we turn our attention to the RPA approach (without appealing to the identity to rCCD). A concept of a \textit{ground state energy} for RPA (which is primarily an excited state theory in quantum chemistry) can be developed by conceptualizing that the RPA excitations are bosonic oscillators and the excitation energies are the corresponding oscillatory frequencies.\cite{rowe2010} From this viewpoint and analogous to the zero point energy of molecular vibrations, the RPA ground state correlation energy for drCCD is
\begin{equation}
E_{drCCD} =E_{dRPA} = \dfrac{1}{2} \sum_{i}(\omega_i - Tr(A)) 
\end{equation}
while for the rCCD the energy is, \begin{equation}
E_{rCCD}=E_{RPA}= \dfrac{1}{4} \sum_{i}(\omega_i - Tr(A))
\end{equation}
In terms of (d)rCCD, we know that $T={Y}{X}^{-1}$, $A + BT = R$ and $R = X{\omega}X^{-1}$. This leads to, 
\begin{equation}
Tr(BT)=Tr(R)-Tr(A)   
\end{equation}
\begin{equation}
Tr(BT)=Tr(X{\omega}X^{-1}) - Tr(A)  
\end{equation}
\begin{equation}
Tr(BT)=\sum_{i}(\omega_i - \sigma_i)  
\end{equation}

where $\sigma_i= Tr(A)$ are configuration interaction singles (CIS) excitation energies. Notwithstanding the debate that RPA is a correlated method, the above equation can be interpreted as an expression for the energy difference between a correlated and a reference uncorrelated method (CIS). Purely from the (d)rCCD view point, we can unambiguously write the correlation energy as $\dfrac{1}{4} Tr(\bar{B}T)$ and $\dfrac{1}{2}Tr(BT)$ for rCCD and drCCD respectively. However, physical attribution of the RPA states to bosonic vibrators leads to the factor $\dfrac{1}{2}$ instead of $\dfrac{1}{4}$ for RPA correlation energy (as far as we know this ambiguity has not been fully resolved).
The RPA or CIS solutions are singlet or triplet states. Both singlet and triplets states enter the correlation energy expression and can be factorized for direct-RPA as,
\begin{equation} 
E_{dRPA} = \dfrac{1}{2} \sum_{i}[{^{\textbf{1}}(\omega_{i} - \sigma_{i} )+{^{\textbf{3}} }}(\omega_{i} - \sigma_{i} )]   
\end{equation}
In the case of RPA, we have two corresponding expressions. One with the spin-flip excitations included is given as, 
\begin{equation}
E_{RPA}^{(spin-flip)} = \dfrac{1}{4} \sum_{i}[{{^{\textbf{1}}(\omega_{i} - \sigma_{i} )} +  {3}( ^{\textbf{3}} (\omega_{i} - \sigma_{i}))}]
\label{spinflipRPA}
\end{equation}
and the other without the spin-flip excitations given as, 
\begin{equation}
E_{RPA}^{(No-spin-flip)} = \dfrac{1}{4} \sum_{i}[{^{\textbf{1}}(\omega_{i} - \sigma_{i} )+{^{\textbf{3}}(\omega_{i} - \sigma_{i} ) }}]
\end{equation}
where superscripts $\textbf{1}$ and $\textbf{3}$ refer to singlets and triplets respectively. There are no spin-flip excitations in dRPA, since all the integrals in $A$ and $B$ matrices are Coulomb only. The spin-flip forms take into account the excitations (and de-excitations) like $a_{\alpha}^{\dagger} i_{\beta}$ and ${i_{\beta}^{\dagger}}{a_{\alpha}}$. For a closed shell ground state, these excitations increase $\hat{s}_z$ by 1 (or decrease by 1 for de-excitations). The operators, $a_{\alpha}^{\dagger} i_{\beta}$, ${i_{\beta}^{\dagger}}{a_{\alpha}}$ and $ \dfrac{1}{\sqrt{2}} (a_{\alpha}^{\dagger} i_{\beta} + {i_{\beta}^{\dagger}}{a_{\alpha}})$ are the triplet occupied-virtual one electron replacement operators corresponding to the $\hat{s}_z = 1, -1$ and $0$ components. When these excitations are included in the RPA problem, the triplet states appear as triply degenerate (we only include the $\hat{s}_z= 0 $ component of the triplet while considering RPA without the spin-flip excitation). Therefore, when the spin-flip excitations are included the RPA correlation energy expression carries a factor 3 for the triplet states reflecting the degeneracy of the state (see Eqn. \ref{spinflipRPA}). 
In order to account for the RPA spin-flip excitations, the rCCD method needs to be reformulated since the canonical rCCD equations do not include spin-flip excitations as a consequence of the antisymmetry of the amplitudes\cite{Klopper2011, Hehn2013, Hehn2015, HEHN2016}. The spin-orbital rCCD equations are given by 
\begin{equation}
{\bra{ab}}{\ket{ij}}+ t_{ij}^{ab} (\epsilon_{a}+\epsilon_b - \epsilon_i - \epsilon_j )+ t_{im}^{ae}{\bra{mb}}{\ket{ej}} + t_{jm}^{be}{\bra{ma}}{\ket{ei}} + t_{im}^{ae}{\bra{mn}}{\ket{ef}} t_{nj}^{bf} = 0
\end{equation}
As opposed to the canonical rCCD, where we need to consider only three unique spin-combinations for the UHF implementations, the spin-flip rCCD has four unique spin combinations corresponding to the  $t_{Ij}^{AB}$, $t_{ij}^{ab}$, $t_{Ij}^{Ab}$ and $t_{Ij}^{aB}$ amplitudes. Equations for the first three combinations are identical to canonical rCCD, and the amplitude equation for $t_{Ij}^{aB} (= t_{i_{\alpha}j_{\beta}}^{a_{\beta}b_{\alpha}})$ is given by,
\begin{equation}
\bra{aB}{\ket{Ij}} + t_{Ij}^{aB} (\epsilon_a + \epsilon_b - \epsilon_i - \epsilon_j) + t_{Im}^{aE}\bra{mB}{\ket{Ej}} +t_{jM}^{Be}
\bra{Ma}{\ket{eI}} = 0
\end{equation}
The spin-orbital energy with the spin-flip excitations is,
\begin{equation}
E_{rCCD}^{spin-flip}=\dfrac{1}{4} Tr(\bar{B}T)
\end{equation}
spin-integrated into the expression for UHF,
\begin{equation}
E_{UrCCD}=\dfrac{1}{4} t_{IJ}^{AB} {\bra{IJ}}{\ket{AB}} + \dfrac{1}{4} t_{ij}^{ab} {\bra{ij}}{\ket{ab}} +\dfrac{1}{2} t_{Ij}^{Ab} {\bra{Ij}}{\ket{Ab}} - \dfrac{1}{2} t_{Ij}^{aB} {\bra{Ij}}{\ket{aB}}
\end{equation}
or spin-adapted RHF,
\begin{equation}
E_{RrCCD}=\dfrac{1}{4} (3(^{3} {B}^{3}{T})+ ^{1}{B}^{1}{T})
\end{equation}
with the singlet and triplet spin-adapted integrals and amplitude defined above (note that $^{3}{T}=t_{Ij}^{aB}$). 

\bibliography{paper_EOM_rCCD}

\end{document}